\documentclass[a4paper]{jpconf}
\usepackage{graphicx,amssymb,amsmath}
\newcommand{\dif}{\mathrm{d}}

\begin{document}

\title{Dynamical freeze-out criterion in event-by-event hydrodynamics}

\author{Hannu Holopainen$^1$ and Pasi Huovinen$^1$}

\address{$^1$ Frankfurt Institute for Advanced Studies, 
Ruth-Moufang-Stra\ss e~1,\\
D-60438 Frankfurt am Main, Germany}

\ead{huovinen@fias.uni-frankfurt.de}

\begin{abstract}
  In hydrodynamical modelling of ultrarelativistic heavy-ion collisions
  the freeze-out is typically assumed to take place on a surface of
  constant temperature or energy density. In this work we apply a
  dynamical freeze-out criterion, which compares the hydrodynamical
  expansion rate with the pion scattering rate, to an event-by-event
  ideal hydrodynamics at the full RHIC collision energy. We present
  hadron spectra and elliptic flow calculated using (2+1)-dimensional
  ideal hydrodynamics, and show the differences between constant
  temperature and dynamical freeze-out criteria. We find that when the
  freeze-out ratio is fixed to one, the different freeze-out criteria
  lead to slightly different spectra and $v_2(p_T)$ in the
  event-by-event calculations.
\end{abstract}

\section{Introduction}

In hydrodynamical modelling of heavy-ion collisions the fluid dynamical
description is usually assumed to break down, and the momentum
distributions to freeze-out, on a thin surface of constant temperature
or energy density. However, it has been argued for a while ago that a
more physical criterion would be when the expansion rate of the fluid
exceeds the scattering rate of the constituents of the
fluid~\cite{Bondorf:1978kz}. Such a dynamical criterion has been
applied to hydrodynamical modelling
before~\cite{Hung:1997du,Eskola:2007zc}, but it has not been widely
used. In this contribution we study how a dynamical freeze-out
criterion affects the final pion and proton distributions and their
elliptic anisotropy in event-by-event hydrodynamics. We extend our
previous results shown in Ref.~\cite{Holopainen:2012id} by using an
equation of state (EoS) for a chemically frozen system.

\section{Fluid dynamics}

Since we want to concentrate on the effects of the freeze-out
criterion, we ignore all the complications of viscosity, and use a
modified version of the ideal fluid model described in
Ref.~\cite{Holopainen:2010gz}. The model is boost invariant, and
includes finite net baryon density, which, however, does not
contribute to pressure. As an EoS we use the $s95p$-PCE-$v1$
parametrisation~\cite{Huovinen:2009yb}, which contains a chemical
freeze-out at temperature $T_\mathrm{chem}=150$ MeV to reproduce the
observed $\pi/p$ ratio at RHIC~\cite{Huovinen:2007xh}.

We initialise the system using a Monte Carlo Glauber approach. The
entropy density is given as a mixture of binary (75\%) and wounded
nucleon (25\%) profiles. A Monte Carlo Glauber model gives only the
positions of the wounded nucleons and binary collisions and we need to
distribute entropy around these positions before we can initialise
hydrodynamical evolution. Our choice is a 2-dimensional Gaussian:
\begin{equation}
    s(x,y) = \textrm{const.} \sum_{\textrm{wn,bc}} \frac{1}{2\pi \Sigma^2} 
             \exp \Big[-\frac{ (x-x_i)^2 + (y - y_i)^2 }{2\Sigma^2} \Big],
\end{equation}
where $\Sigma$ is a free parameter controlling the width of the
Gaussian. Typical values for this fluctuation size parameter is order
of $0.5$~fm, and our choice here is $0.8$~fm (see
Ref.~\cite{Holopainen:2012id}). For net-baryon density a pure wounded
nucleon profile is used. To obtain averaged initial state, we evaluate
an average of 1000 event-by-event fluctuating initial states, which
include the above mentioned smearing by Gaussians.  As initial time of
the evolution we use $\tau_0 = 0.6$~fm.

\section{Dynamical freeze-out criterion}

A usual requirement for the validity of fluid dynamics is that the
Knudsen number must be much larger than one. Since the ratio of the
expansion rate $\theta$ to the scattering rate $\Gamma$ is the ratio
of an inverse of a macroscopic scale to an inverse of a microscopic
scale, it can be identified as a Knudsen number. Thus the requirement
that the scattering rate is larger than the expansion rate is equivalent
to requiring that the Knudsen number is larger than one. We assume
that fluid dynamics is valid until $K = \theta/\Gamma = 1$, and use
$K_\mathrm{f}=1$ as the freeze-out criterion in the dynamical
case. As a baseline for comparison we use freeze-out at constant
temperature of $T_\mathrm{f}=120$ MeV.

The expansion rate of the system, $\theta = \partial_\mu u^\mu$, can
be evaluated once the flow velocity is known, but the scattering rate
is not given by hydrodynamics, but requires microscopic
calculations. Several evaluations of the scattering rate of pions in
equilibrated pion gas can be found in the
literature~\cite{Hung:1997du,Eskola:2007zc,Daghigh:2001gy}, but since
we are interested in pion scatterings in chemically frozen hadron gas,
we calculate the rate ourselves. In semiclassical approximation the
scattering rate of pions in kinetically (but not necessarily
chemically) equilibrated hadron-resonance gas is given by
\begin{equation}
    \Gamma = \frac{1}{n_\pi(T,\mu_\pi)} \sum_i \int \dif^3 p_\pi \dif^3 p_i\,
      f_\pi(T,\mu_\pi) f_i(T,\mu_i)\, v_{\pi i}(s)\, 
      \sigma_{\pi i} (s),
\end{equation}
where $n_\pi$ is the density of pions, $f_i(T,\mu_i)$ is the thermal
distribution function, $v_{\pi i}$ is the relative velocity,
$\sigma_{\pi i}$ is the cross section for scattering of pion with
particle $i$, and the summation over $i$ runs over all particle
species included in the EoS. Cross sections are evaluated as in the
UrQMD hadron cascade~\cite{Bass:1998ca}, \emph{i.e.}\ the main
contribution comes from resonance formation which is evaluated using
the relativistic Breit-Wigner formula. The scattering rates are
different for different hadrons, and therefore different hadrons
should freeze-out at different times. Implementation of such a
sequential freeze-out in fluid dynamics is very difficult, and it has
also been argued that a proton requires more than one scattering with
pions to change its momentum significantly~\cite{Prakash:1993bt}. To
simplify the problem we thus assume that all hadrons freeze-out
simultaneously.

Once the constant temperature or constant Knudsen number surface is
found, the particle and resonance distributions are calculated using
the conventional Cooper-Frye prescription, and the contributions from
resonance decays are added to particle distributions.

\section{Results}

\begin{figure}[t]
 \begin{minipage}{18pc}
  \includegraphics[width=18pc]{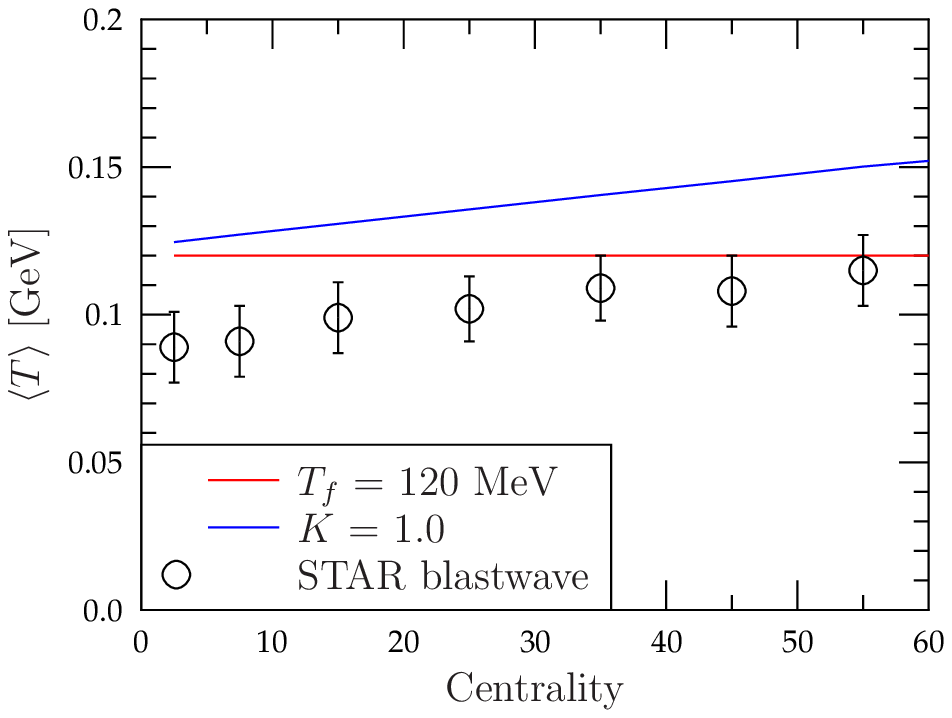}
 \end{minipage}\hspace{2pc}%
 \begin{minipage}{18pc}
  \includegraphics[width=18pc]{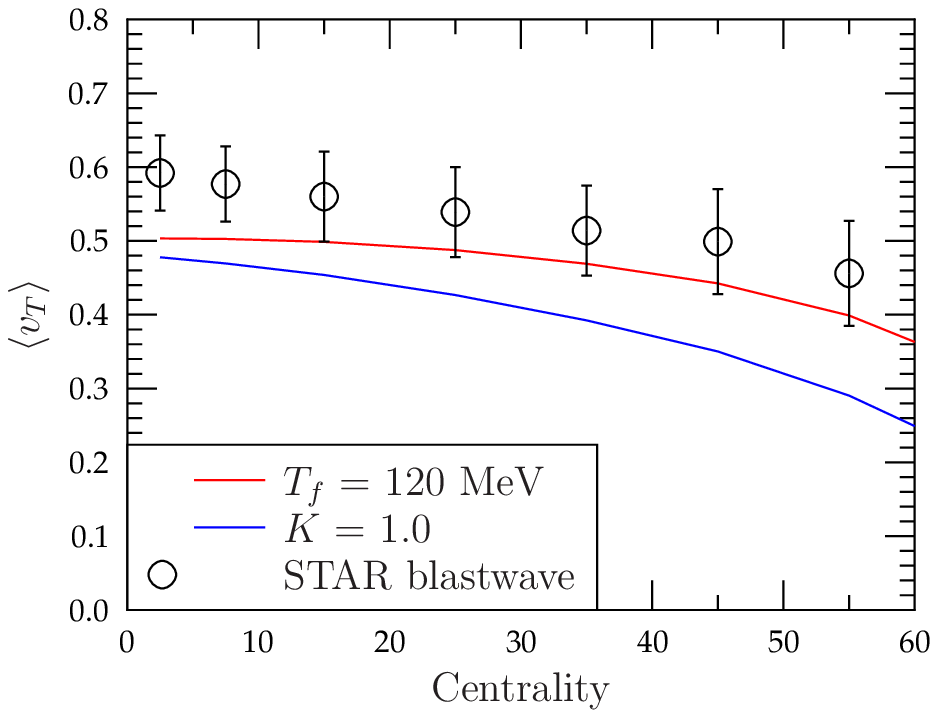}
 \end{minipage}
\caption{\label{ave}Average temperature (left) and flow velocity
  (right) on the freeze-out surface as function of centrality in
  $Au+Au$ collisions at $\sqrt{s_\mathrm{NN}}=200$ GeV using
  freeze-out at constant temperature (red) or at constant Knudsen
  number (blue). Ideal fluid calculations are compared with the STAR
  blast-wave fit~\cite{Adams:2003xp}.}
\end{figure}

We consider Au+Au collisions at Relativistic Heavy Ion Collider (RHIC)
with $\sqrt{s_{NN}} = 200$~GeV. In Fig.~\ref{ave} we show the average
temperature and flow velocity on the freeze-out surface as function of
centrality. The initial state is the average one, and we compare our
results to those of the blast-wave fit by the STAR
collaboration~\cite{Adams:2003xp}. Trivially constant temperature
freeze-out leads to a centrality independent temperature, whereas
constant Knudsen number leads to the hotter surface the more
peripheral the collision. As well known, the more central the
collision, the larger the flow velocity, but the dependence on
centrality is stronger for the dynamical freeze-out criterion. For
both criteria the temperature and velocity differs from the blast-wave
fits, but this is understandable: The shape of the freeze-out surface
affects the favoured values of temperature and flow velocity, and
fluid dynamical calculation leads to more complicated shape of the
freeze-out surface than a simple blast-wave fit. What is interesting
although, is that the slopes of the centrality dependence of
temperature and flow velocity are similar in the blast-wave fit and in
the calculation with dynamical freeze-out criterion.

\begin{figure}[h]
\includegraphics[width=18pc]{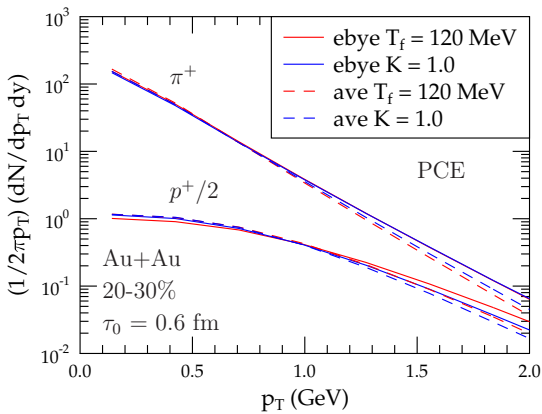}\hspace{2pc}%
\begin{minipage}[b]{16pc}\caption{\label{label}The $p_T$-distributions
    of positive pions and protons in $Au+Au$ collisions at
    $\sqrt{s_\mathrm{NN}}=200$ GeV using freeze-out at constant
    temperature (red) or at constant Knudsen number
    (blue). Calculations are done for both averaged (dashed) and
    fluctuating (solid) initial states. 

\raisebox{5ex}{ }}
\end{minipage}
\end{figure}

Fig.~\ref{label} depicts the $p_T$-spectra of positive pions and
protons at $20-30$\% centrality. As well
known~\cite{Holopainen:2010gz}, when all the other parameters are kept
unchanged, event-by-event calculation leads to stiffer spectra than
the averaged initial state calculation. The constant Knudsen number
freeze-out leads to slightly steeper spectra than the constant
temperature freeze-out, except for pions in the event-by-event
case. Thus, to obtain similar $p_T$-distributions slightly larger
Knudsen number $K_\mathrm{f} > 1$ might be required for the averaged
initial state. On the other hand, since the pion spectrum is
independent of the freeze-out criterion in the event-by-event case, a
slightly different initial time, or initial shape might be required to
reproduce the $T_\mathrm{f} = 120$ MeV result in that case.

\begin{figure}[t]
\begin{minipage}{18pc}
\includegraphics[width=18pc]{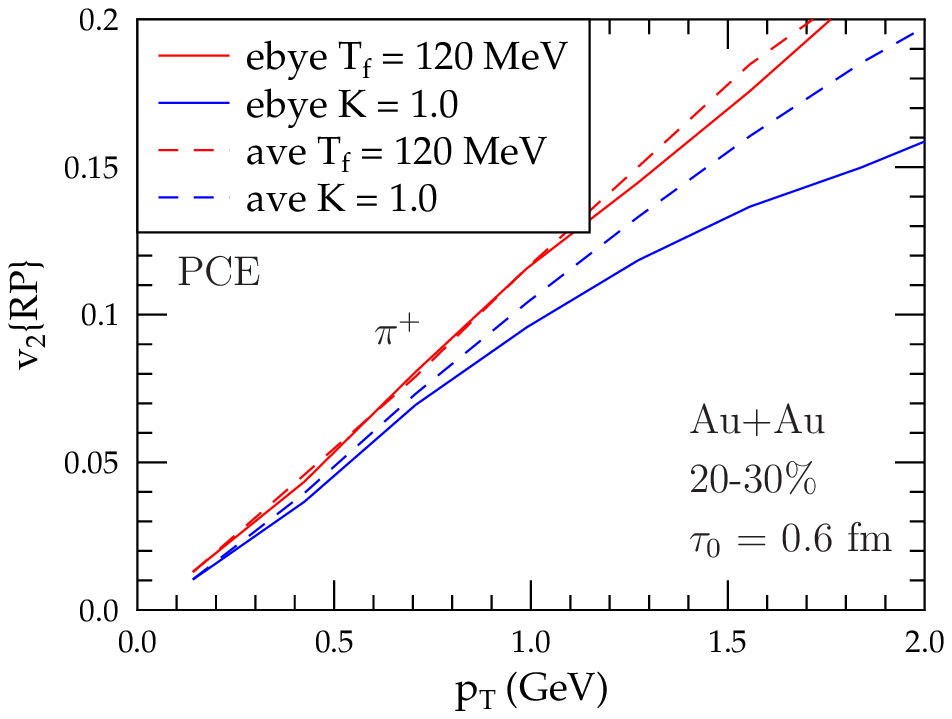}
\end{minipage}\hspace{2pc}%
\begin{minipage}{18pc}
\includegraphics[width=18pc]{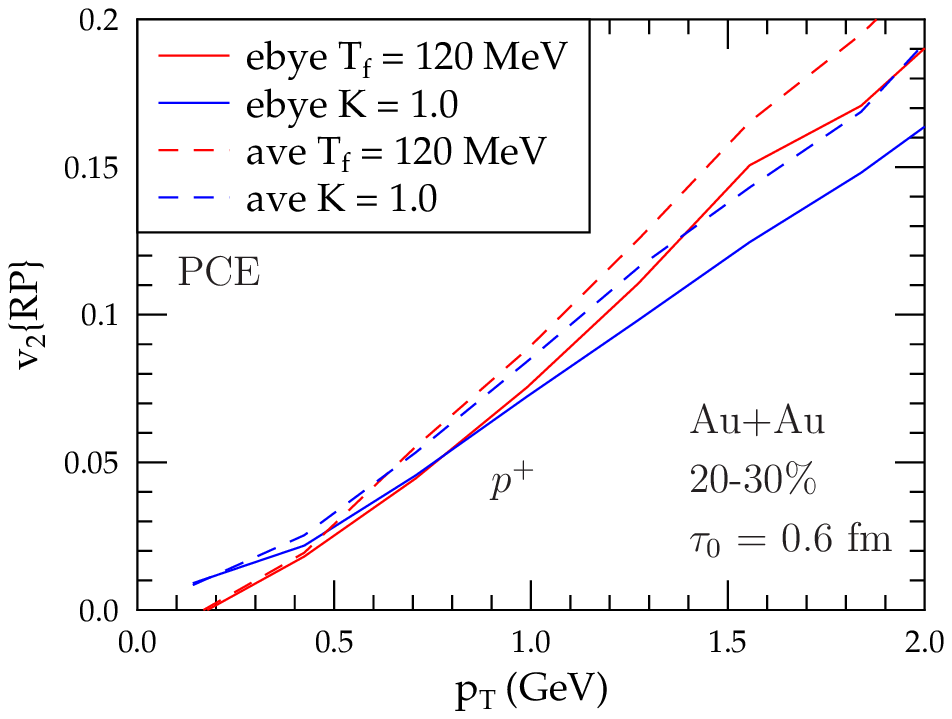}
\end{minipage}
\caption{\label{v2}The $p_T$-differential elliptic flow of pions
  (left) and protons (right) in $Au+Au$ collisions at
  $\sqrt{s_\mathrm{NN}}=200$ GeV using averaged (dashed) or
  fluctuating (solid) initial states, and freeze-out at constant
  $T = 120$ MeV temperature (red) or at constant Knudsen number $K =
  1$ (blue).}
\end{figure}

In Fig.~\ref{v2} the pion and proton $p_T$-differential $v_2$ are shown
in collisions with $20-30$\% centrality. Since our $p_T$-distributions
were not tuned to be the same in all the cases, one cannot draw any
strong conclusions based on the $v_2(p_T)$ results shown
here. Nevertheless, when constant temperature freeze-out is used, the
pion $v_2(p_T)$ is not sensitive to the event-by-event fluctuations of
the initial state~\cite{Holopainen:2010gz}, but the dynamical
criterion creates some sensitivity to fluctuations at $p_T\gtrsim 1$
GeV. More strikingly, the dynamical criterion reduces pion $v_2(p_T)$
by $\sim 10$\% both in the average initial state and in the
event-by-event calculation. For protons the situation is reversed: The
fluctuations have now a clear reducing effect independent of the
freeze-out criterion, whereas the freeze-out criterion has a weaker
and $p_T$-dependent effect. At very low $p_T$ the dynamical criterion
increases and at large $p_T$ decreases the proton $v_2(p_T)$.

\section{Conclusions}

We argued that the constant temperature/density freeze-out is an
oversimplification, but the effects of the freeze-out criterion on
spectra and $v_2(p_T)$ turned out to be small. The $\sim 10$\% reduction
in pion $v_2(p_T)$ is interesting, but it is unknown whether it
survives if $K_\mathrm{f}$ is taken as a free parameter, and chosen to
lead to the same $p_T$-distributions than $T_\mathrm{f}$.

\ack
The work of H.H.\ was supported by the Extreme Matter Institute (EMMI)
and the work of P.H.\ by BMBF under contract no.\ 06FY9092.

\end{document}